\documentstyle[preprint,aps]{revtex}
\begin{document}
\tightenlines 

\title{Evidence for Supercurrent Quantization in Interfacial Josephson 
Junctions} \author{M.J. Black, B.W. Alphenaar\cite{arthur}, and H. 
Ahmed} \address{University of Cambridge Microelectronics Research 
Centre and Hitachi Cambridge Laboratory, Cavendish Laboratory, 
Madingley Road, Cambridge, CB3 0HE, UK} \date{\today}

\maketitle

\begin{abstract}

We observe a series of discontinuities in the above gap 
current/voltage characteristics of highly transmissive Nb/Si 
superconductor/semiconductor interfaces.  The bias at which each 
switch occurs decreases in quantized steps as a function of 
temperature and magnetic field.  The observed step-size is the same 
for both dependencies.  We propose that the steps correspond to 
supercurrent quantization through granular Josephson junctions formed 
near the interface.

\end{abstract}

\pagebreak

\narrowtext


The conductance through a quantum point contact in a two-dimensional 
electron gas (2DEG) is quantized to $\rm 2e^{2}/h$ times the number of 
transmitted one-dimensional channels\cite{QPC}.  Analogous behavior is 
predicted for a small area Josephson junction formed by a narrow 
constriction between two superconductors\cite{SPC}.  In this case, 
quantization is in the critical current, and is given by $\rm 
e\Delta_{0}/\hbar$ times the number of transmitted one-dimensional 
channels, where $\rm \Delta_{0}$ is the pairing potential of the 
superconductor.  Recently, Takayanagi et al.\ \cite{Tak} fabricated 
the first working superconducting quantum point contacts, using Nb 
contacts to a submicron width InAs 2DEG, modulated by a split-gate.  
They observed steps in both the critical current and the normal 
conductance as a function of gate bias.  However, substantial 
fluctuations were observed in the step height, and the critical 
current was not quantized at the theoretically predicted values.  
Muller et al.\ \cite{Mul1} measured the conductance and critical 
current through a break junction of Nb.  They observed steps in the 
critical current consistent with the theoretically predicted value, 
suggesting that individual conductance channels are gradually shut off 
as the filament is broken.  More recent experiments\cite{Mul2} imply, 
however, that such critical current steps might be caused by abrupt 
atomic rearrangements, which change the normal resistance\cite{Grib}.

This paper presents intriguing new evidence for the quantization of 
the critical current in small area Josephson junctions.  We perform 
detailed measurements of the interface resistance of Nb/Si junctions.  
Below a transition temperature of 1.5 K, we observe a series of 
switches in the above gap DC current/voltage characteristics.  As the 
temperature increases towards the transition, or as a magnetic field 
is applied, the switching bias decreases in a series of quantized 
steps.  The observed step-size is equal in both temperature and 
magnetic field dependencies.  We propose that the switches occur as we 
exceed the critical supercurrent of Josephson junctions formed between 
$\rm{Nb_{3}Si}$ particles and the Nb contact.  By varying the temperature 
or magnetic field we are able to reduce the junction area, and observe 
a corresponding quantized change in the critical current.

Fig.\ 1(a) shows a schematic drawing of our Nb/Si junctions.  A Nb 
contact is made to a 53 nm thick single crystal n$^{++}$ Si layer 
which has been wafer bonded to 100 nm of SiO$_{2}$ on a p$^{+}$ Si 
substrate.  2.0 K Hall measurements of the Si layer indicate that the 
carrier density $n$ = 5.3 $\times$ 10$^{25}$m$^{-3}$ and the elastic 
scattering length $l_{e}$ = 10.8 nm.  We create a highly transmissive 
interface by first sputter-etching the Si surface at 200 W for 10 
minutes in an Ar pressure of 3 $\times$ 10$^{-2}$ mbar.  This removes 
a 25 $\mu$m $\times$ 25 $\mu$m square of the (100) Si top-layer, and 
exposes clean (111) Si side-surfaces.  Without breaking the vacuum we 
then deposit 180 nm of Nb, resulting in a total Nb/Si contact area of 
approximately $A_{c}$ = 6 ($\mu$m)$^{2}$.  We measure the critical 
temperature of our Nb to be $T_{c}$ = 9.0 K, indicating a high quality 
film.  Four-terminal measurements of the Nb/Si junction resistance are 
made via Ag contacts to the Si, and Au contacts to the Nb.  The 
effects we describe below have been observed in four different 
devices, each fabricated using the same method.

In the inset to Fig.\ 2, the differential resistance (dV/dI) of a 
typical junction is plotted as a function of applied bias.  
Measurements at three temperatures (1.6, 1.2 and 0.35 K) are shown, 
normalized with respect to the resistance at high bias.  In each 
trace, there is a broad dip in dV/dI corresponding to the Nb gap 
energy ($\Delta_{0} = \pm$1.6 meV).  Further measurements (not shown) 
demonstrate that the dip survives up to the Nb transition temperature.  
This dV/dI feature has been reported by a number of 
groups\cite{hes,bigdip} and can be attributed to Andreev reflection at 
the Nb/Si interface\cite{Andy,BTK}.  In the 0.35 K trace we also 
observe a strong sub-gap resistance dip.  (This first appears at 
temperatures below 1.1 K.) A similar feature was reported by Heslinga 
et al.\ \cite{hes} and attributed to the appearance of a `proximity 
induced gap,' in the Si layer.  More simply, the RSJ (resistively 
shunted junction) model predicts that a sub-gap dip can be observed if 
a Josephson junction forms in parallel with the interface 
resistance\cite{tink}.

In our case, a possible explanation for the sub-gap resistance dip is 
suggested by the temperature dependence of dV/dI at zero bias (not 
normalized), plotted in the main part of Fig.\ 2.  The differential 
resistance decreases sharply at a temperature $T_{1}$ = 1.5 K, from 
890 to 480 $\Omega$.  This drop, which strongly resembles a 
superconducting transition, occurs at the reported transition 
temperature of $\rm{Nb_{3}Si}$ \cite{Tc}.  If we assume that the ratio 
between the transition temperature $T_{1}$ and the gap energy 
$\Delta_{Nb_{3}Si}$ equals that for Nb, then $\Delta_{Nb_{3}Si}$ = 
0.25 meV. This is similar to the observed resistance dip width of 
$\pm$0.28 mV at {\em T} = 0.35 K. These results imply that the sub-gap 
feature is due to Andreev reflection from $\rm{Nb_{3}Si}$ particles 
that form near the interface during fabrication.  $\rm{Nb_{3}Si}$ has 
a high eutectic temperature (645$^{\circ}$ C), but it has been shown 
that Nb diffuses nonuniformly into Si at our maximum process 
temperature of 300$^{\circ}$C (immediately following the sputter-etch), 
and forms small intermixed regions \cite{silicide}.  We also measured 
junctions made without the sputter-etch, and these showed no sub-gap 
dip or resistance transition.

Figure\ 3(a) gives results of a four-terminal DC measurement of the 
Nb/Si interface (see inset) at 1.6 and 0.35 K, corresponding to 
temperatures above and below the transition temperature, $T_{1}$.  At 
{\em T} = 1.6 K, the I-V curve has a weak Schottky barrier dependence.  
At {\em T} = 0.35 K, the current rises above the 1.6 K value, as 
expected from the results of Fig.\ 2.  As the applied bias increases, 
the measured current and voltage switch discontinuously due to a 
sudden increase in the interface resistance.  (A switch is observed in 
both the current and the voltage across the interface $V_{m}$ because 
only the total applied bias $V_{a}$ is fixed at each measurement 
point.)  A total of six switches are observed with increasing bias, 
until the above and below transition I-V characteristics are almost 
identical.  In Fig.\ 3(a), we also plot the 0.35 K characteristic 
measured with a transverse magnetic field of {\em B} = 1.0 T (dashed line).  
The I-V is almost identical to the above transition trace measured at 
0 T---only a small excess current is observed at low bias.  We note 
that these results are completely reproducible, and that 
non-reproducible noise due to heating is not observed until the 
current $>$ 200 $\mu$A. The maximum power through the device is $\sim$1 
$\mu$W, while our pumped $^{3}$He cryosat is able to maintain a 100 
$\mu$W heat load at 0.35 K for 3.5 hours.

In Fig.\ 3(b), we plot a series of 26 I-V characteristics for 
temperatures between 1.6 and 0.35 K. The lowest lying trace 
corresponds to {\em T} = 1.6 K. The traces above this correspond to 
the temperatures 1.55 K, 1.50 K,\ldots, 0.35 K and are offset by 0.25 
$\mu$A, 0.50 $\mu$A,\ldots, 6.25 $\mu$A, respectively.  Plotted in 
this way, the resistance switches reveal a remarkable temperature 
dependence.  As the temperature increases from 0.35 K, the bias at 
which each switch occurs decreases discontinuously, in a series of 
steps.  For any particular switch, the step width is approximately 
constant.  If we consider the three most clearly defined switches 
(marked $S_{1}$, $S_{2}$, and $S_{3}$), the average width of the first 
four steps is $dV_{1}$ = 1.60 mV, $dV_{2}$ = 1.82 mV and $dV_{3}$ = 
2.04 mV, with a maximum variation (per switch) of 3 \%.  The steps can 
be seen more clearly if the switching bias $V_{n}$ is plotted as a 
function of temperature.  This is shown in the inset of Fig.\ 3(b) where 
$V_{n}$ is plotted for the three switches $S_{1}$, $S_{2}$ and $S_{3}$ 
as a function of the reduced temperature $T/T_{1}$.  The biases are 
normalized by the average step widths, $dV_{n}$.  In each case, three 
or more equally spaced steps are observed.  As the temperature 
approaches $T_{1}$, the switches merge together.

Further measurements provide more convincing evidence for the 
quantization of the switching bias.  Figure 4(a) shows the normalized 
switching bias as a function of temperature taken from observations of 
the lowest bias switch in a second device.  These points are compiled 
from a set of 94 I-V measurements (similar to those shown in Fig.\ 
3(b)) taken at equally spaced temperatures between 1.30 and 0.37 K. A 
total of 12 well-defined steps are observed, each containing at least 
three bias points.  The average step width, {\em dV}, is 0.42 mV. We 
point out that 4 other switches were observed in this device, and each 
showed a similar temperature dependence, although {\em dV} is 
different for each switch.

We also studied the influence of magnetic field on the switching bias.  
Figure 4(b) shows the normalized switching bias for the switch shown in 
Fig.\ 4(a) as a function of magnetic field perpendicular to the current 
path.  These points were compiled from a set of 130 I-V measurements 
taken at magnetic fields equally spaced between 0 and 0.33 Tesla ({\em 
T} = 0.37 K).  The data clearly show that the switching bias is also 
quantized as a function of magnetic field.  We observe 12 well-defined 
steps each containing 6 or more {\em B}-field points.  The bias is 
again normalized using {\em dV} = 0.42 mV. For {\em B} $>$ 0.25 T, the 
steps dissapear, and the switching bias decreases contiuously with 
magnetic field.  Most remarkably, a comparison between Figs.\ 4(a) and (b) 
shows that the quantized step-size is equal to the same value in the 
{\em B}-dependence as in the {\em T}-dependence.  The background {\em 
T} and {\em B}-dependences, however, are clearly different.  Similar 
results were observed in all 5 switches of this device.

To understand these results, we now consider the Nb/Si interface in 
more detail (see Fig.\ 1(b)).  Current flows across the interface by 
electron tunneling through the Si depletion region.  Because of the 
high doping density, the average inter-dopant spacing ($n^{-1/3} 
\approx$ 2.7 nm) is comparable to the depletion width (3.7 
nm)\cite{depl}.  In this case, it has been shown that the depletion 
width varies substantially across the contact and significant 
transmission occurs only at the most transparent sections of the 
interface \cite{grainy}.  We calculate the approximate area of the 
transparent regions as follows.  The average transmission coefficient 
per quantum channel is given by $t_{av} = R_{S}/R_{j}$ where $R_{S}$ 
is the Sharvin resistance and $R_{j}$ is the measured junction 
resistance.  From Fig.\ 2, $R_{j}$ at 1.6 K = 890 $\Omega$, and 
$R_{S}$ for our contact is 20 m$\Omega$ so that $t_{av}$ = 2.3 
$\times$ 10$^{-5}$.  On the other hand, a fit of the normalized 
dV/dI dip at 1.6 K to the BTK formalism gives a transmission 
coefficient for the transparent regions of $t_{A}$ = 0.84\cite{BTK}.  
The maximum area of the transparent regions is then approximately 
$A_{T} = A_{c} \times (t_{av}/t_{A})$ = 164 nm$^{2}$, where $A_{c}$ is the 
total contact area.  Below the transition, at {\em T} = 1.2 K, the 
resistance drops to $R_{j}$ = 480 $\Omega$ , but the size of the 
normalized dip in dV/dI remains roughly the same (see inset, 
Fig.\ 2).  This means that $t_{A}$ stays constant, while the 
transmissive area increases to approximately \linebreak 164 nm$^{2}$ 
$\times$ (890 $\Omega$ / 480 $\Omega$) = 304 nm$^{2}$.

Additional conduction pathways below $T_{1}$ are created if 
$\rm{Nb_{3}Si}$ particles form near the interface during sputtering.  
The pathlength for tunneling via the particles is reduced at the 
$\rm{Nb_{3}Si}$ transition temperature, and the effective transmissive 
area increases.  The cross-sectional area for the $\rm{Nb_{3}Si}$ 
particles is approximately equal to the increase in the transmissive 
area or $\approx$ 140 nm$^{2}$.  As shown in Fig.\ 1(b), if the normal 
coherence length $\xi_{n}$ reaches the $\rm{Nb_{3}Si}$ particles a 
supercurrent can flow through weak-link Josephson junctions between 
the $\rm{Nb_{3}Si}$ particles and the Nb ($J_{1}$ and $J_{2}$ in Fig.\ 
1(b)).

Consider now the switches observed in the I-V charactersitics of Fig.  
3.  The switches progressively transform the $T<T_{1}$ resistance 
(0.35 K) into the $T>T_{1}$ resistance (1.6 K) with increasing bias.  
Each switch is thus a partial breakdown of the conduction pathways 
added at $T_{1}$.  This could be possibly explained by assuming that 
Joule heating raises the temperature of the conduction pathways above 
$T_{1}$.  In this case\cite{Skocpol}, the switching bias should follow 
a $(T_{1} - T)^{1/2}$ dependence.  Our observed switching bias, 
however, increases more slowly at low {\em T} than this.  In addition, 
it is unclear how a heating model can explain the quantization we 
observe in Fig.\ 4.

Resistance steps have also been reported by Hebard and Vandenberg in 
the I-V characteristics of high-resistance granular lead 
films\cite{pb}.  These were later explained as the switching of rows 
of underdamped Josephson junctions formed between grains, lying 
perpendicular to the current path\cite{zant}.  A switch occurs as a 
row moves from a zero-voltage state, to one in which a vortex 
undergoes a phase-slip process.  The switches we observe might 
possibly be due to a similar process ocurring among granular Josephson 
junction arrays within the depletion region.  Typically, though, the 
separation between switches in a voltage biased Josephson-junction 
array are independent of small magnetic field, and as the temperature 
is increased, the switches simply 
disappear\cite{zant,yu}---quantization is not observed.

We might assume, then, that a group of Josephson junction pathways are 
associated with each particular switch.  As the coherence length in 
the depletion region decreases with increasing {\em T} or {\em B}, 
individual conduction channels are cut off, resulting in discrete 
steps.  But since the observed step spacing is quantized for a 
particular switch, this would require an identical contribution from 
each random pathway.

We instead propose that the switches are due to the critical current 
transition of {\em individual} nanometer scale Josephson junctions 
formed between $\rm{Nb_{3}Si}$ particles and the Nb contact.  Consider 
a simple circuit model for the Nb/Si contact (Fig.\ 1(c)).  Transport 
through the nth silicide particle is given by a Josephson junction 
$J_{n}$ in series with a $\rm{Nb_{3}Si}$/Si interface resistance 
$R_{n}$.  In the superconducting state, the Josephson junction 
resistance is zero, and the current through $J_{n}$ is simply $I_{n} = 
V / R_{n}$.  The switching bias is thus proportional to the current 
carried by the Josephson junction at the critical current transition.

If the area of the Josephson junction is small enough, the switching 
bias should be quantized to $V = Ne\Delta_{0}/R_{n}\hbar$, where N is the 
number of one-dimensional transmitted channels, and 
$e\Delta_{0}/R_{n}\hbar$ is equal to the quantized step size $dV$.  
In our case, $N = k_{F}^{2}A/4\pi$, where {\em A} is the 
cross-sectional area of the junction.  To produce the observed 
quantization, we assume that {\em A} decreases as a function of 
{\em B} and {\em T}.  Since the normal coherence length $\xi_{n}$ 
decreases with increasing {\em B} and {\em T}, this is reasonable.  
The number of modes {\em N} in each junction (Fig.\ 3 and 4)
is simply equal to the normalized switching bias $V/dV$.  From the 
maximum number of modes, we can calculate the approximate 
cross-sectional area $A_{n}$ of the particles corresponding to the 
three switches $S_{1}$, $S_{2}$, and $S_{3}$ in Fig.\ 3.  This gives 
$A_{1}$ = 56 nm$^{2}$, $A_{2}$ = 84 nm$^{2}$ and $A_{3}$ = 131 
nm$^{2}$, roughly corresponding to the cross-sectional area determined 
from our simple estimate above (140 nm$^{2}$), and giving an average 
particle radius on the order of $\lambda_{F}$.

We expect that, in the limit of large {\em N}, the {\em T} and {\em B} 
dependences in Fig.\ 4 should be described by $I_{c}(T, B)$ for a 
large area point contact.  The envelope of our quantized {\em T}-dependence 
appears similar to $I_{c}(T)$ for a large area device\cite{Kulik}, 
while the envelope of the quantized {\em B}-dependence resembles the low 
field behavior of a screened Josephson junction\cite{tink}.  The field 
required to insert a single flux quantum in such small junctions 
($\sim$ 100 nm$^{2}$) is larger than the critical field of Nb, so that no 
oscillatory dependence is expected.  A detailed understanding would 
require a rigorous analysis of $I_{c}(T, B)$ in the crossover between 
macroscopic and nano-scale junctions.

In conclusion, we observe evidence that the maximum supercurrent 
through Josephson junction pathways formed near a Nb/Si interface is 
quantized, in agreement with theoretical predictions.  This could 
provide a new and straightforward way to study supercurrent transport 
in nanometer-scale Josephson junctions.

The authors thank F. Hekking, T. Klapwijk, M. de Jong, and D. Williams 
for valuable discussions and S. Newcomb for providing TEM analysis.

\begin{figure}

\vspace{1cm}

\caption{(a) Schematic drawing of our Nb/Si junctions.  (b) A 
representative region of the Nb/Si interface.  $\rm{Nb_{3}Si}$ 
particles, which form during Nb deposition, provide additional 
tunneling pathways.  At low enough temperatures the normal coherence 
length $\xi_{n}$ extends to create Josephson junctions $J_{1}$ and 
$J_{2}$.  (c) Circuit model of the Nb/Si interface.  The 
$\rm{Nb_{3}Si}$ Josephson junctions $J_{n}$ have normal state 
resistances $R^{\prime}_{n}$ and lie in series with the 
$\rm{Nb_{3}Si}$/Si interface resistances $R_{n}$.  The in-parallel 
Schottky barrier resistance is $R$.}

\vspace{1cm}

\caption{Zero-bias differential resistance of a highly transmissive 
Nb/Si junction verses temperature.  $T_{1}$ is the $\rm{Nb_{3}Si}$ 
transition temperature.  {\bf Inset}: The normalized differential resistance 
verses bias at 1.6, 1.2 and 0.35 K.  The 1.2 and 1.6 K traces are 
offset by 0.1 and 0.2, respectively. The gap widths for Nb and 
$\rm{Nb_{3}Si}$ are indicated.} 

\vspace{1cm}

\caption{(a) Four-terminal measurement of the DC current and voltage 
across the Nb/Si interface at {\em T} = 0.35 K and {\em T} = 1.6 K. 
Each trace shows measurements made at 1000 equally spaced values of 
the applied bias $V_{a}$ (see inset).  The dashed line shows the 
characteristics for {\em B} = 1.0 T and {\em T} = 0.35 K. (b) DC current and 
voltage characteristics plotted for 26 equally spaced temperatures 
between 0.35 K and 1.6 K. The traces are offset for clarity. {\bf 
Inset} The normalized switching bias as a function of reduced 
temperature for the three switches indicated.}

\vspace{1cm}

\caption{ The normalized switching bias as a function of (a) reduced 
temperature and (b) magnetic field for a typical resistance switch.} 

\end{figure}


\begin{references}
\bibitem[*]{arthur} To whom correspondence should be 
addressed.
\end{references}
\end{document}